\documentclass[11pt,a4paper]{article}

\usepackage[a4paper,margin=1in]{geometry}
\usepackage{graphicx}
\usepackage{amsmath,amssymb}
\usepackage{url}
\usepackage{cite}
\usepackage[hidelinks]{hyperref}
\usepackage[font=small,labelfont=bf,justification=centering]{caption}

\begin{document}

\renewcommand{\thefootnote}{\fnsymbol{footnote}}

\begin{center}
{\Large\bfseries Mobility Heterogeneity in a 2D Gaussian Lattice Polymer: A Dynamic Monte Carlo Study\par}
\vspace{0.6em}
Arpan Dey $^{1,}$\footnotemark[1]\\
$^1$ Universit\'e de Montpellier, Montpellier, France\\
Email: \texttt{arpand2004@gmail.com}
\end{center}

\footnotetext[1]{Corresponding author.}

\renewcommand{\thefootnote}{\arabic{footnote}}

\begin{abstract}
We study mobility heterogeneity in a two-dimensional Gaussian lattice polymer using dynamic Monte Carlo simulations. The polymer dynamics is generated from a local three-monomer move dictionary, which explicitly enumerates allowed bond-preserving updates on a square lattice. As a homogeneous benchmark, this dictionary reproduces the expected Rouse-like behavior of an ideal chain, including the crossover in monomer mean-squared displacement (MSD) and the center-of-mass diffusion scaling $D_{\rm cm} \sim N^{-1}$. We then introduce a two-block version of the model in which the two halves of the chain are updated with different attempt rates, $\omega_A$ and $\omega_B$, while the local move dictionary remains unchanged. For $\rho=\omega_A/\omega_B>1$, the more frequently updated block shows a larger block-resolved MSD at early and intermediate times, producing a positive normalized MSD asymmetry. However, numerical measurements show that the center-of-mass diffusion coefficient remains consistent with $D_{\rm cm} \sim N^{-1}$ for all rate ratios studied. We invoke a simple coarse-grained Rouse argument to explain this result analytically. In this minimal Gaussian setting, rate-induced mobility heterogeneity modifies internal relaxation without changing the Rouse scaling of center-of-mass transport.
\end{abstract}

\section{Introduction}

Polymer dynamics is governed by both the stochastic motion of individual monomers and the connectivity constraints imposed by the chain backbone. Even in the absence of explicit long-range interactions, a local displacement of one monomer is transmitted through neighboring bonds and contributes to collective relaxation. The Rouse model provides the standard minimal description of this process, treating a polymer as an overdamped chain of beads connected by harmonic springs and driven by thermal noise~\cite{rouse1953}.

Dynamic Monte Carlo methods on lattices provide a complementary coarse-grained setting in which polymer connectivity and local relaxation can be implemented through discrete bond-preserving moves. Classical local-move schemes, including end-bond rotations and kink jumps, have long been used to model polymer dynamics on lattices~\cite{verdier1962,baschnagel2004,manka2013}. These models are useful because the microscopic update rules are explicit: one can directly specify which local moves are allowed and how often different monomers are selected for attempted updates.

Standard lattice polymer simulations often use a single local update rule for the whole chain. However, polymer dynamics can become heterogeneous when different segments experience different effective mobilities, local constraints, temperatures or active forcing. Related questions have been studied in heterogeneous Rouse models, active Rouse chains, and partially active polymers, where segmental mobility or local driving modifies chain-level relaxation and transport~\cite{hung2018,osmanovic2017,osmanovic2018,vatin2024}.

Here, we study a rate-heterogeneous two-block Gaussian lattice polymer in two dimensions. The chain evolves through local bond-preserving Monte Carlo moves, encoded in a three-monomer move dictionary. We first use the homogeneous Gaussian chain as a benchmark and verify that the dictionary reproduces the expected Rouse-like MSD crossover and the scaling $D_{\rm cm} \sim N^{-1}$. We then divide the chain into two equal blocks and assign different attempt rates, $\omega_A$ and $\omega_B$, while keeping the allowed local moves fixed throughout the chain. Thus, mobility heterogeneity is implemented purely through update frequency, without changing the local move geometry or adding forces or energetic interactions. The resulting dynamics is analyzed using block-resolved MSDs, a normalized MSD asymmetry, and the chain-length dependence of the center-of-mass diffusion coefficient. Finally, we use a coarse-grained Rouse argument to show analytically why the rate contrast changes the diffusion prefactor but preserves the scaling $D_{\rm cm} \sim N^{-1}$.

\section{Rouse reference and lattice dynamics}

\subsection{The Rouse model}

The Rouse model~\cite{rouse1953} provides the simplest dynamical reference for a flexible polymer in a viscous medium. A polymer is represented as a chain of $N$ monomers connected by harmonic springs, with each monomer undergoing overdamped stochastic motion due to friction and thermal noise. Let $\mathbf{r}_i(t)$ denote the position of the $i$-th monomer. For an interior monomer, the continuum Rouse equation may be written as:
\begin{equation}
\zeta \frac{d\mathbf{r}_i}{dt}
= k(\mathbf{r}_{i+1}+\mathbf{r}_{i-1}-2\mathbf{r}_i)+\boldsymbol{\eta}_i(t),
\end{equation}
where $\zeta$ is the friction coefficient, $k$ is the spring constant, and $\boldsymbol{\eta}_i(t)$ represents thermal noise. The spring term couples each monomer to its nearest neighbors along the chain, so that even purely local fluctuations generate collective relaxation.

For an ideal Gaussian chain, where self-avoidance is neglected and monomers are allowed to overlap, the typical polymer size scales as $R \sim aN^{0.5}$, where $a$ is the characteristic bond length. The motion of the center of mass is diffusive at long times, but its diffusion coefficient decreases with chain length as:
\begin{equation}
D_{\rm cm} \sim N^{-1}.
\end{equation}
This scaling follows from the fact that the total friction of the chain grows linearly with the number of monomers, meaning longer chains diffuse more slowly as whole objects, even though their internal modes continue to fluctuate.

The longest relaxation time of the chain, usually called the Rouse time, scales as:
\begin{equation}
\tau_R \sim N^2.
\end{equation}
This timescale separates local monomer relaxation from global chain diffusion. At intermediate times, before the whole chain has relaxed, individual monomers show constrained subdiffusive motion, which is commonly characterized through the mean-squared displacement (MSD). At times larger than $\tau_R$, the polymer moves diffusively through its center of mass. These standard scalings provide the homogeneous benchmark against which the heterogeneous model introduced later will be compared.

\subsection{Lattice Monte Carlo analogue of Rouse dynamics}

We model the polymer as an ordered chain of $N$ monomers on a 2D square lattice. Consecutive monomers are constrained to occupy nearest-neighbor sites, so the bond length is fixed at $a=1$. In the present work, we focus on the Gaussian case, where monomer overlap is allowed. Connectivity is imposed through discrete bond-preserving Monte Carlo moves, rather than through harmonic springs as in the continuum Rouse model.

\begin{figure}[!htbp]
    \centering
    \includegraphics[width=0.50\textwidth]{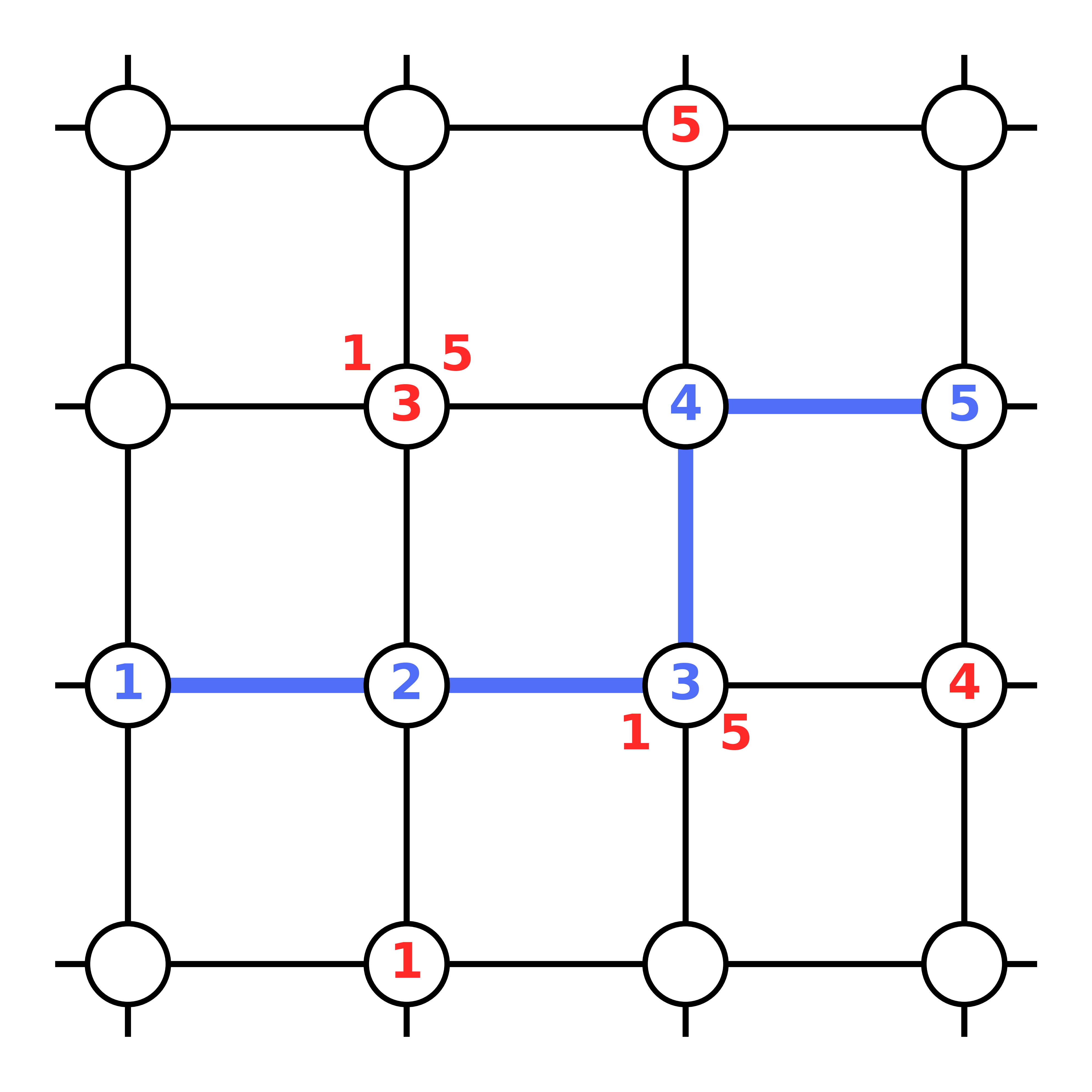}
    \caption{Local bond-preserving moves on a 2D square lattice for the Gaussian polymer. The chain runs through monomers 1, 2, 3, 4, 5. End monomers can move to any neighboring lattice site that preserves their single bond. Internal monomers must preserve bonds to both neighbors. Since the present model is restricted to two dimensions, crankshaft moves are not used here~\cite{verdier1962}.}
    \label{fig:localmoves}
\end{figure}

At each update, one monomer is chosen at random and moved only if the proposed position preserves the required bond lengths. Figure~\ref{fig:localmoves} illustrates the local rules. The chain runs from monomer 1 through 2, 3, 4 to 5. Since monomer 1 is bonded only to monomer 2, it can move to any of the three red sites marked ``1,'' while keeping the bond length fixed. The same applies to monomer 5, which is bonded only to monomer 4. These are end moves.

Internal monomers are more constrained because they are bonded to two neighbors. Monomer 2, lying between monomers 1 and 3 on a straight segment, has no allowed move. By contrast, monomers 3 and 4 sit at corners and can move to the diagonally opposite corner of the local square while preserving both adjacent bonds. These are the usual kink jumps, or corner flips, of lattice polymer dynamics.

One Monte Carlo sweep consists of $N$ attempted updates, so that each monomer is selected once on average per sweep. In the homogeneous benchmark all monomers follow the same rules and are selected with equal probability. We use this case to check that the lattice dynamics reproduces the expected Rouse-like results.

\subsection{Three-monomer move dictionary}

In a local lattice update, the motion of an internal monomer is fully determined by the positions of its two bonded neighbors. Thus, instead of recomputing the allowed positions for all the monomers from scratch at every update, we encode the local geometry in a three-monomer move dictionary. For an internal monomer $i$, the relevant local configuration is the triplet $(i-1,i,i+1)$. A proposed move of monomer $i$ is allowed only if the new position remains at unit lattice distance from both neighboring monomers. The two end monomers are treated separately, since they have only one bonded neighbor.

\begin{figure}[!htbp]
    \centering
    \includegraphics[width=0.96\textwidth]{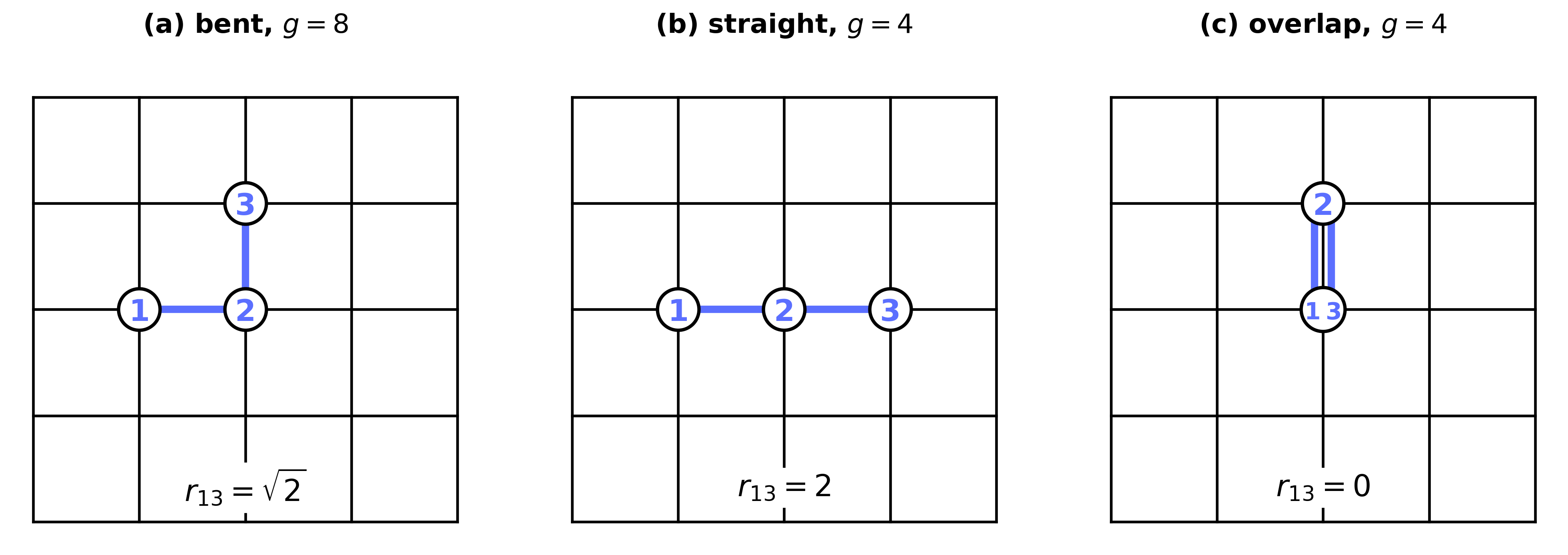}
    \caption{Three local generators for a Gaussian three-monomer chain on a 2D square lattice. The bent generator has end-to-end distance $r_{13}=\sqrt{2}$ and degeneracy $g=8$; the straight generator has $r_{13}=2$ and $g=4$; and the overlapping generator has $r_{13}=0$ and $g=4$. Together these give sixteen elementary configurations used in the three-monomer move dictionary.}
    \label{fig:generators}
\end{figure}

On a 2D square lattice, a Gaussian three-monomer chain has three possible local generators, shown in Fig.~\ref{fig:generators}. In the bent generator, the end-to-end distance is $r_{13}=\sqrt{2}$; in the straight generator, $r_{13}=2$; and in the overlapping generator, $r_{13}=0$, where monomers 1 and 3 occupy the same site. Including orientations and ordering of the chain, these generators produce $g=8$, $g=4$, and $g=4$ local configurations, respectively, giving a total of sixteen elementary configurations, which form the three-monomer move dictionary used throughout the lattice simulations.

Each internal monomer of the polymer locally behaves as the middle monomer of such a three-monomer system. The full chain dynamics can therefore be generated by applying this dictionary to each internal monomer, while retaining the separate end-move rule from Fig.~\ref{fig:localmoves} for the end monomers. This dictionary approach is computationally efficient because the allowed internal moves are pre-tabulated rather than recomputed at every update, and this gives a common implementation for both the homogeneous benchmark and the rate-heterogeneous two-block model introduced later. This simplification is especially natural in two dimensions, where the Gaussian three-monomer chain has only sixteen elementary configurations; an explicit enumeration of all local configurations would become much less practical in higher dimensions.

\subsection{Homogeneous Gaussian benchmark}

Before introducing mobility heterogeneity, it is necessary to benchmark this discrete lattice algorithm against the continuum Rouse model. We simulate a homogeneous Gaussian chain, where all monomers obey identical move rules and multiple monomers are permitted to occupy the same lattice site. As a reminder, in the classical Rouse model for an ideal Gaussian chain, the center-of-mass diffusion coefficient scales as $\sim N^{-1}$, and the longest characteristic relaxation time, the Rouse time, scales as $\sim N^2$.

To simulate the dynamics on the 2D lattice, the polymer is initialized as a chain of monomers with fixed bond lengths. The initial chain is generated as a random walk starting from the center of the lattice. Each subsequent monomer is placed on a randomly chosen neighboring lattice site, while remaining within the simulation box. All simulations are performed on a $500\times 500$ square lattice. Since the largest chain length studied here is $N=100$, the lattice is much larger than the polymer size, and boundary effects are negligible over the simulated time windows.

In a single Monte Carlo attempt, a monomer is selected uniformly at random. If it is an end monomer, an end-bond rotation is proposed; if it is an internal monomer, a move is proposed using the pre-tabulated local move dictionary. The proposed move is accepted if the target site preserves the requisite bond lengths to adjacent monomers. Time is measured in Monte Carlo sweeps, where one sweep consists of $N$ independent update attempts, ensuring that each monomer is selected once on average per sweep.

\begin{figure}[!htbp]
    \centering
    \includegraphics[width=0.76\textwidth]{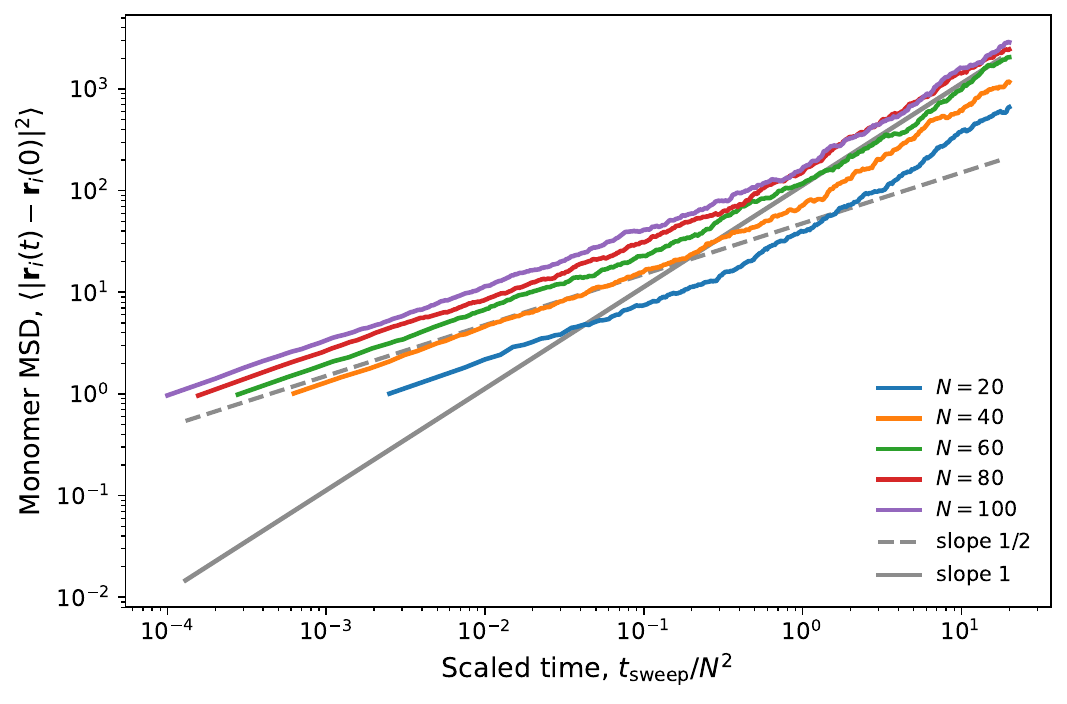}
    \caption{Log-log plot of monomer mean-squared displacement for homogeneous Gaussian lattice polymers simulated using the three-monomer move dictionary. Time is measured in Monte Carlo sweeps, with one sweep corresponding to $N$ attempted monomer updates, and is scaled as $t_{\rm sweep}/N^2$. Results are shown for $N=20,40,60,80$, and $100$, with each curve averaged over 100 independent runs. Simulations were run up to $t_{\rm sweep}/N^2=20$. The dashed and solid gray guide lines indicate slopes $1/2$ and $1$, respectively. The crossover occurs at $t_{\rm sweep}/N^2=O(1)$, consistent with Rouse-like relaxation up to a prefactor.}
    \label{fig:homomsd}
\end{figure}

We measure the monomer mean-squared displacement, $\mathrm{MSD}(t)=\langle |\mathbf{r}_i(t)-\mathbf{r}_i(0)|^2\rangle$, averaged over all monomers and over independent simulation runs, and plot it on log-log axes as a function of the scaled time $t_{\rm sweep}/N^2$. Figure~\ref{fig:homomsd} shows the ensemble-averaged monomer MSD for homogeneous Gaussian chains of different lengths. The curves exhibit the expected Rouse-like crossover: at smaller times, monomer motion is constrained by chain connectivity and follows an approximately subdiffusive regime, while at longer times the polymer crosses over to ordinary diffusion through its center-of-mass motion. This confirms that the dictionary-based lattice dynamics reproduces the expected Rouse relaxation timescale. The crossover does not occur exactly at $t_{\rm sweep}/N^2=1$, but at a value of the same order. This is expected because the Rouse scaling $\tau_R \propto N^2$ holds up to a numerical prefactor.

\begin{figure}[!htbp]
    \centering
    \includegraphics[width=0.72\textwidth]{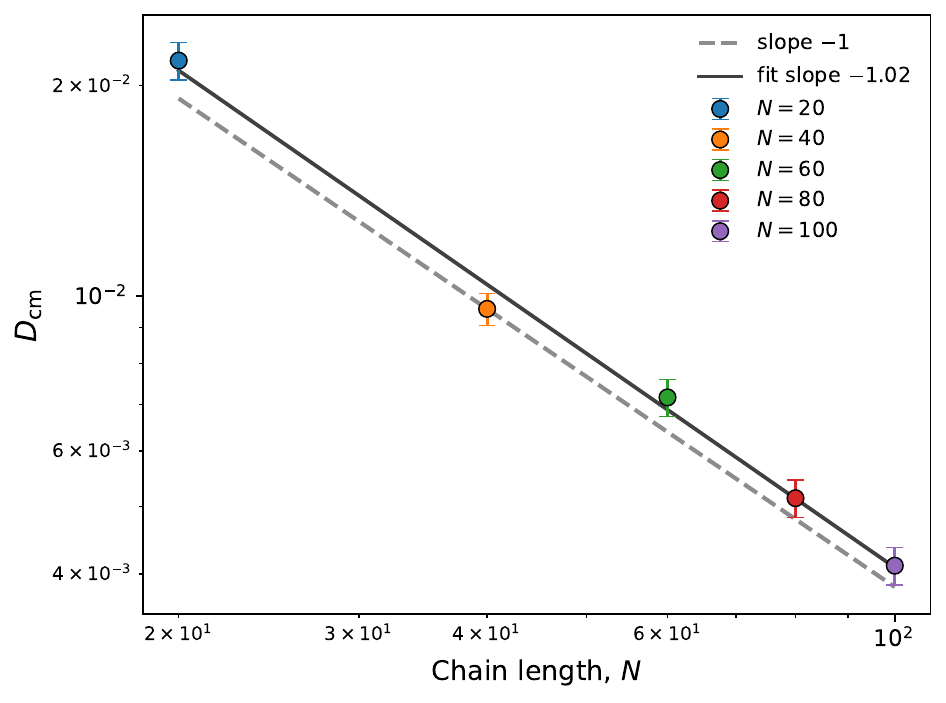}
    \caption{Center-of-mass diffusion coefficient for homogeneous Gaussian lattice polymers simulated using the three-monomer move dictionary. Results are shown for $N=20,40,60,80$, and $100$, with error bars estimated from 100 independent runs. For each chain length, $D_{\rm cm}$ is extracted from the time-averaged center-of-mass MSD using $\mathrm{MSD}_{\rm cm}(\Delta t)\simeq 4D_{\rm cm}\Delta t$, over the lag-time window $0.1\leq \Delta t_{\rm sweep}/N^2\leq 10$. The log-log plot shows the expected Rouse scaling $D_{\rm cm}\sim N^{-1}$ as a dashed gray line, while the solid line shows the fitted scaling of the simulation data.}
    \label{fig:homodcm}
\end{figure}

As a second check, we measure the chain-length dependence of the center-of-mass diffusion coefficient. For each trajectory, the center-of-mass position $\mathbf{R}_{\rm cm}(t)$ is recorded, and the center-of-mass mean-squared displacement is computed as:
\begin{equation}
\mathrm{MSD}_{\rm cm}(\Delta t)
= \left\langle |\mathbf{R}_{\rm cm}(t+\Delta t)-\mathbf{R}_{\rm cm}(t)|^2\right\rangle_t,
\end{equation}
where $\langle \cdots \rangle_t$ denotes averaging over all available starting times $t$ along the trajectory. In two dimensions, diffusive center-of-mass motion gives:
\begin{equation}
\mathrm{MSD}_{\rm cm}(\Delta t) \simeq 4D_{\rm cm}\Delta t.
\end{equation}
We therefore extract $D_{\rm cm}$ from the linear growth of $\mathrm{MSD}_{\rm cm}(\Delta t)$, using the fitting window $0.1\leq \Delta t_{\rm sweep}/N^2 \leq 10$. Figure~\ref{fig:homodcm} shows $D_{\rm cm}$ as a function of $N$ on log-log axes. The data follow $D_{\rm cm}\sim N^{-1}$, consistent with the Rouse prediction for an ideal chain.

\section{A rate-heterogeneous two-block model}

\subsection{Model definitions}

\begin{figure}[!htbp]
    \centering
    \includegraphics[width=0.55\textwidth]{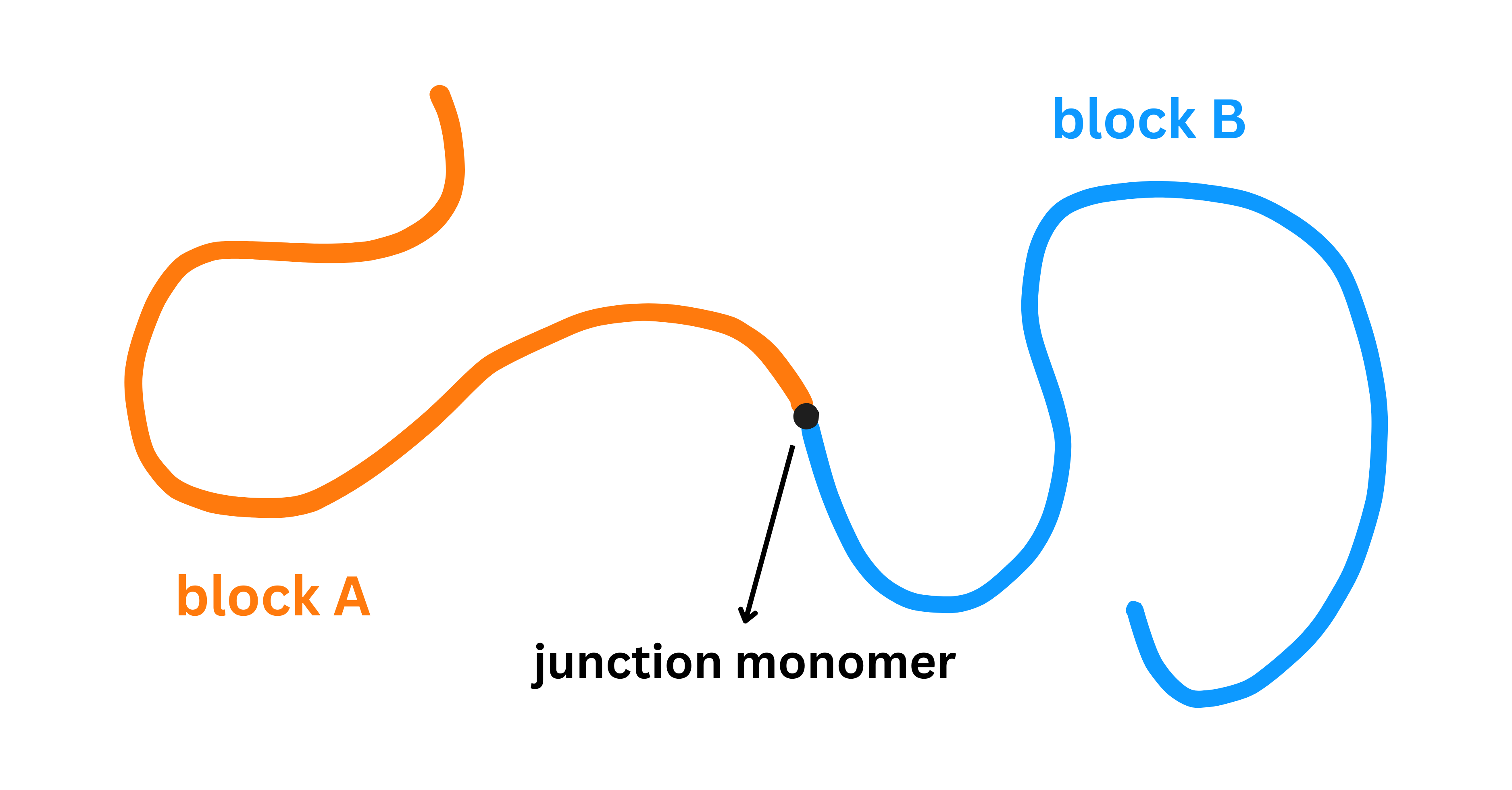}
    \caption{Schematic of the two-block model. The first half of the chain is defined as block A and the second half as block B. The two blocks obey the same bond-preserving local move dictionary, but their monomers are selected for attempted updates with different rates $\omega_A$ and $\omega_B$. The black dot marks the junction monomer at the interface between the two blocks.}
    \label{fig:twoblock}
\end{figure}

We now introduce a rate-heterogeneous two-block model built from the homogeneous Gaussian lattice benchmark. The polymer of length $N$ is divided into two equal contiguous blocks: the first $N/2$ monomers form block A, shown in orange in Fig.~\ref{fig:twoblock}, and the remaining monomers form block B, shown in blue. The interface between the two blocks is marked by the junction monomer, shown in black.

Both blocks use the same local three-monomer move dictionary described above. Thus, the allowed bond-preserving moves are not changed from one block to the other; the heterogeneity is in how often monomers from each block are selected for attempted updates. We assign the two blocks attempt rates $\omega_A$ and $\omega_B$, and define the rate ratio:
\begin{equation}
\rho=\frac{\omega_A}{\omega_B}.
\end{equation}

In the simulations, this rate ratio is implemented through biased monomer selection. For each Monte Carlo attempt, the probability of choosing block A is proportional to $\omega_A N_A$, while the probability of choosing block B is proportional to $\omega_B N_B$. Since we use equal block sizes, $N_A=N_B=N/2$, we have:
\begin{equation}
P_A=\frac{\rho}{\rho+1},\qquad
P_B=\frac{1}{\rho+1}.
\end{equation}

Once a block has been chosen, one monomer is selected uniformly from that block and updated using the three-monomer move dictionary. 

The homogeneous case $\rho=1$ corresponds to the reference Rouse-like dynamics, since both blocks are selected with equal attempt rates. We then study $\rho=2$ and $\rho=4$ as controlled departures from this homogeneous limit, where block A is updated more frequently than block B. The model realizes mobility heterogeneity through block-dependent attempt rates only, without adding forces, energetic interactions, or block-dependent geometrical rules, which allows us to isolate whether rate heterogeneity primarily affects internal block-resolved relaxation or also modifies the center-of-mass transport of the full chain.

To quantify the relative mobility of the two blocks, we also define a normalized MSD asymmetry:
\begin{equation}
\mathcal{A}_{\rm MSD}(t)
=\frac{\mathrm{MSD}_A(t)-\mathrm{MSD}_B(t)}{\mathrm{MSD}_A(t)+\mathrm{MSD}_B(t)},
\end{equation}
where $\mathrm{MSD}_A(t)$ and $\mathrm{MSD}_B(t)$ denote the monomer-averaged mean-squared displacements of block A and block B, respectively. Since both MSDs are non-negative, $\mathcal{A}_{\rm MSD}(t)$ is bounded between $-1$ and $+1$. A value of $\mathcal{A}_{\rm MSD}(t)$ close to zero corresponds to nearly symmetric block motion, while positive values indicate that block A is more mobile than block B.

\subsection{Dynamics of the Gaussian two-block model}

\begin{figure}[!htbp]
    \centering
    \includegraphics[width=1.0\textwidth, keepaspectratio]{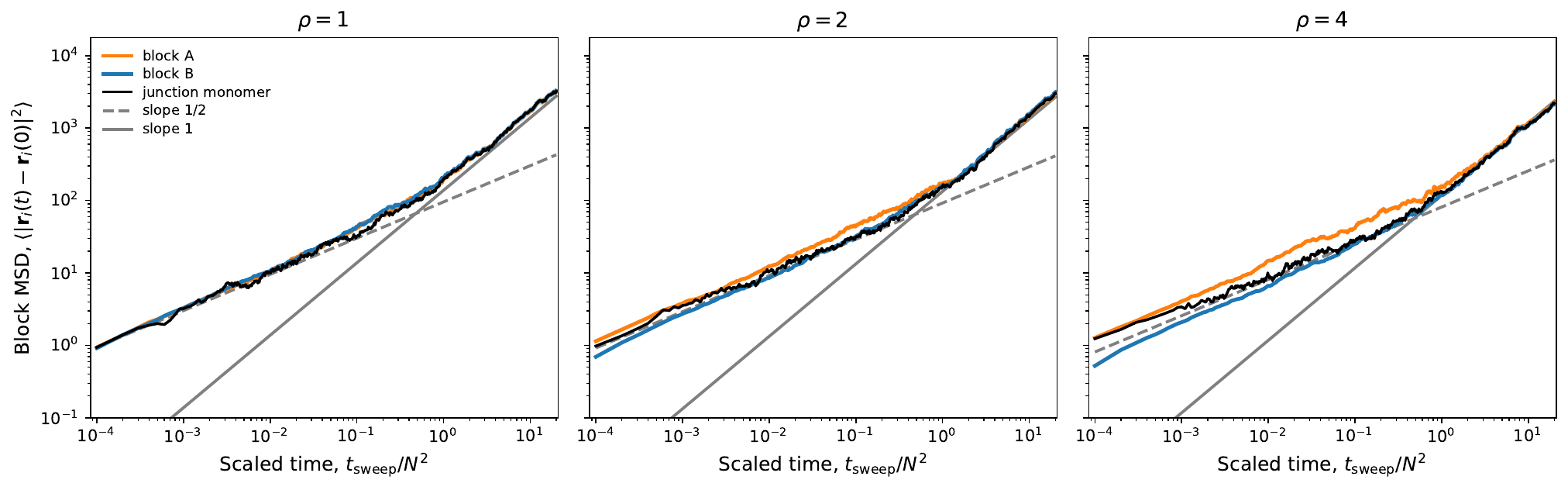}
    \caption{Log-log plot of block-resolved MSDs for the Gaussian two-block polymer with $N=100$. Results are shown for $\rho=1,2$, and $4$, with each curve averaged over 100 independent runs up to $t_{\rm sweep}/N^2=20$. Orange and blue curves denote the monomer-averaged MSDs of blocks A and B, respectively, while the black curve denotes the MSD of the junction monomer. The dashed and solid gray guide lines indicate slopes $1/2$ and $1$.}
    \label{fig:blockmsd}
\end{figure}

\begin{figure}[!htbp]
    \centering
    \includegraphics[width=0.76\textwidth]{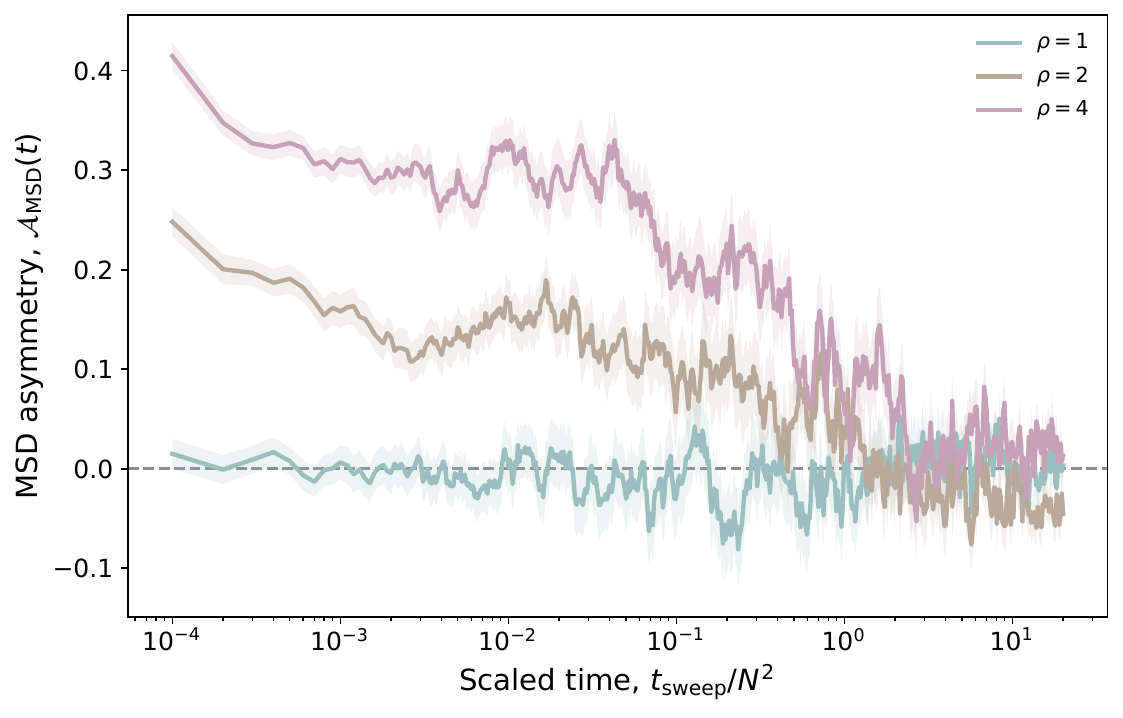}
    \caption{Time dependence of $\mathcal{A}_{\rm MSD}(t)$ for the Gaussian two-block polymer with $N=100$. Results are shown for $\rho=1,2$, and $4$, averaged over 100 independent runs. The homogeneous case $\rho=1$ fluctuates around zero, while $\rho=2$ and $\rho=4$ show positive asymmetry at early and intermediate times, reflecting the larger mobility of block A.}
    \label{fig:asymmetry}
\end{figure}

\begin{figure}[!htbp]
    \centering
    \includegraphics[width=0.78\textwidth]{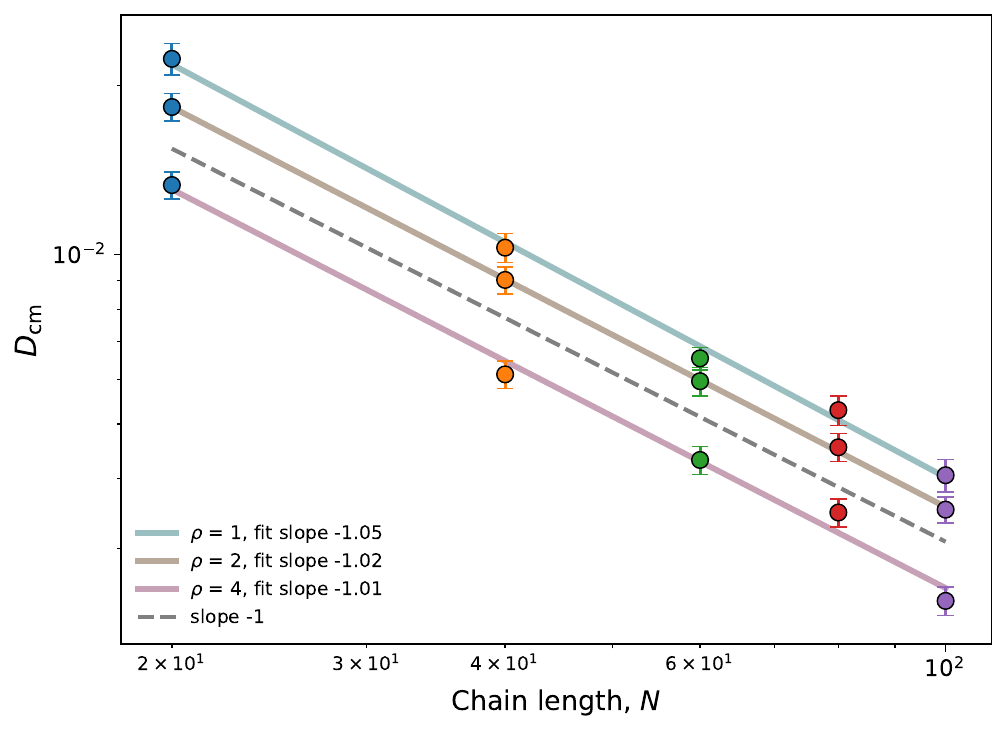}
    \caption{Log-log plot of the center-of-mass diffusion coefficient $D_{\rm cm}$ as a function of chain length for the Gaussian two-block polymer. Results are shown for $N=20,40,60,80$, and $100$, and $\rho=1,2$, and $4$, averaged over 100 independent runs. For each $(N,\rho)$, $D_{\rm cm}$ is extracted from the time-averaged center-of-mass MSD using $\mathrm{MSD}_{\rm cm}(\Delta t)\simeq 4D_{\rm cm}\Delta t$, over the lag-time window $0.1\leq \Delta t_{\rm sweep}/N^2\leq 10$. The fitted slopes remain close to $-1$, consistent with the Rouse scaling $D_{\rm cm}\sim N^{-1}$.}
    \label{fig:blockdcm}
\end{figure}

We simulate the Gaussian two-block polymer, where overlap between monomers is allowed. In the plots shown in Figs.~\ref{fig:blockmsd}, \ref{fig:asymmetry} and \ref{fig:blockdcm}, the trajectories were evolved up to $t_{\rm sweep}/N^2=20$ and averaged over 100 independent runs. Figure~\ref{fig:blockmsd} shows the block-resolved MSDs for $\rho=1,2$, and $4$. For $\rho=1$, the two blocks are statistically equivalent and the MSDs of block A, block B, and the junction monomer overlap. For $\rho>1$, the MSD of block A lies systematically above that of block B over a broad range of times, reflecting the larger update rate of block A.

All three MSDs retain the qualitative Rouse-like crossover structure seen in the homogeneous benchmark; at early and intermediate scaled times, the curves follow a subdiffusive trend close to the slope $1/2$ guide line, while at later times they cross over toward ordinary diffusive growth with slope $1$. Thus, changing the relative attempt rates produces a visible block-level mobility imbalance, but still preserves the Rouse-like relaxation pattern of the Gaussian chain.

Figure~\ref{fig:asymmetry} shows the time series of $\mathcal{A}_{\rm MSD}(t)$ for the above time window. For $\rho=1$, the asymmetry fluctuates around zero, as expected for the homogeneous case. For $\rho=2$ and $\rho=4$, the asymmetry is positive over early and intermediate times, with the larger rate contrast producing the larger asymmetry. At longer times, the curves decrease toward zero. This indicates that rate heterogeneity strongly affects internal motion, but becomes less significant once the chain enters collective long-time motion. The residual fluctuations near late times are expected to decrease further with larger ensembles and longer trajectories.

Finally, Figure~\ref{fig:blockdcm} shows the center-of-mass diffusion coefficient as a function of chain length for $\rho=1,2$, and $4$, using $N=20,40,60,80$, and $100$, with results averaged over 100 independent runs. The center-of-mass MSD is diffusive over the measured range, and $D_{\rm cm}$ was extracted from the lag-time interval $0.1\leq \Delta t_{\rm sweep}/N^2\leq 10$. This interval avoids early and late-time effects, and using a wider time window gives the same scaling within numerical scatter. The fitted exponents are close to $-1$ for all three rate ratios, showing that the Gaussian two-block model preserves the Rouse scaling $D_{\rm cm}\sim N^{-1}$. 

\subsection{Analytical argument for the $D_{\rm cm}\sim N^{-1}$ scaling in the two-block model}

The numerical result in Fig.~\ref{fig:blockdcm} can be understood from a simple coarse-grained Rouse argument. In the simulations, the two blocks are assigned attempt rates $\omega_A$ and $\omega_B$, with rate ratio $\rho=\omega_A/\omega_B$. At each Monte Carlo attempt, block A is chosen with probability
\begin{equation*}
P_A=\frac{\omega_A}{\omega_A+\omega_B}
\end{equation*}
and block B is chosen with probability
\begin{equation*}
P_B=\frac{\omega_B}{\omega_A+\omega_B}.
\end{equation*}
Since one Monte Carlo sweep consists of $N$ attempted updates, these probabilities determine how the fixed number of attempted moves is redistributed between the two blocks. For two equal blocks, $N_A=N_B=N/2$, the actual per-monomer attempt rates per sweep are therefore:
\begin{equation}
q_A=\frac{NP_A}{N/2}
=\frac{2\omega_A}{\omega_A+\omega_B}
=\frac{2\rho}{\rho+1},
\qquad
q_B=\frac{NP_B}{N/2}
=\frac{2\omega_B}{\omega_A+\omega_B}
=\frac{2}{\rho+1}.
\end{equation}
Thus $q_A/q_B=\rho$; and $\rho>1$ means monomers in block A are attempted more frequently than monomers in block B, while the total number of attempted updates per sweep remains fixed at $N$.

A monomer that is updated more frequently has a larger single-monomer diffusivity. In a continuum Rouse description, this corresponds to a smaller effective friction. Thus, at the coarse-grained level, we identify:
\begin{equation}
\gamma_i \propto \frac{1}{q_i},
\end{equation}
where $q_i=q_A$ for monomers in block A and $q_i=q_B$ for monomers in block B.

With this correspondence, a heterogeneous Rouse equation may be written as:
\begin{equation}
\gamma_i\frac{d\mathbf{r}_i}{dt}
= k(\mathbf{r}_{i+1}+\mathbf{r}_{i-1}-2\mathbf{r}_i)+\boldsymbol{\eta}_i(t),
\end{equation}
where $\gamma_i$ is the effective friction of monomer $i$. Summing over all monomers, the internal spring forces cancel because every spring exerts equal and opposite forces on its neighboring monomers. Thus, we have:
\begin{equation}
\sum_{i=1}^N \gamma_i\frac{d\mathbf{r}_i}{dt}
= \sum_{i=1}^N \boldsymbol{\eta}_i(t).
\end{equation}
At times longer than the internal relaxation time, $t\gg \tau_R$, the chain translates collectively, so that
\begin{equation}
\frac{d\mathbf{r}_i}{dt}\simeq \frac{d\mathbf{R}_{\rm cm}}{dt}
\end{equation}
for all monomers.

At long times, the summed equation therefore has the same form as the overdamped equation for a single diffusing object, with an effective friction $\Gamma_{\rm tot}$. Since the diffusion coefficient of an overdamped Brownian object is inversely proportional to its friction, the center-of-mass diffusion scales as:
\begin{equation}
D_{\rm cm}\propto \frac{1}{\Gamma_{\rm tot}}
=\frac{1}{\sum_i \gamma_i}.
\end{equation}
For the two-block polymer with $N_A=N_B=N/2$,
\begin{equation*}
\sum_i \gamma_i \propto \frac{N}{2q_A}+\frac{N}{2q_B}.
\end{equation*}
Substituting this in $D_{\rm cm}\propto 1/\sum_i\gamma_i$ gives:
\begin{equation}
D_{\rm cm}\propto
\frac{1}{\dfrac{N}{2q_A}+\dfrac{N}{2q_B}}
=\frac{4\rho}{N(\rho+1)^2}.
\label{eq:dcm-prefactor}
\end{equation}
Thus, even though the rate contrast changes the prefactor of $D_{\rm cm}$, the $D_{\rm cm}\sim N^{-1}$ scaling is preserved. For $\rho=1$, Eq.~\eqref{eq:dcm-prefactor} reduces to the homogeneous result $D_{\rm cm}\propto N^{-1}$. For $\rho>1$, the unequal update rates increase the total effective friction of the chain relative to the homogeneous case, producing a lower prefactor. This accounts for the downward shift of the $D_{\rm cm}$-versus-$N$ curves in Fig.~\ref{fig:blockdcm}, while their fitted slopes remain close to $-1$.

The same argument also extends to polymers with more than two blocks. As long as each block occupies a fixed fraction of the chain and the local move rules are unchanged, the center-of-mass diffusion coefficient keeps the same scaling $D_{\rm cm}\sim N^{-1}$.

\section{Discussion and outlook}

The results above show that the rate-heterogeneous two-block model, simulated using the three-monomer move dictionary, remains Rouse-like in the Gaussian limit. When block-dependent attempt rates are introduced, the internal dynamics becomes asymmetric: the more frequently updated block shows a larger MSD at early and intermediate times, and this is captured by the positive MSD asymmetry for $\rho>1$. However, this internal mobility imbalance does not change the leading center-of-mass scaling. Within numerical accuracy, all three rate ratios studied remain consistent with $D_{\rm cm}\sim N^{-1}$. We justify this result analytically, and further show that this result is not specific to two blocks: chains with more general rate-heterogeneous block partitions should retain the same $D_{\rm cm}\sim N^{-1}$ scaling as long as the fraction of monomers in each block remains fixed and the local move geometry is kept uniform.

We focus here on the Gaussian version of the model because the local three-monomer dictionary gives a clean and controlled lattice analogue of Rouse-like dynamics in that setting. Extending the same construction to self-avoiding chains is not automatic. A simple implementation using the three-monomer dictionary in which moves to occupied sites are rejected mixes several effects: excluded volume, reduced acceptance of local moves, finite-time relaxation, and the specific restrictions of the local dictionary~\cite{madras1987}. Preliminary tests with self-avoidance, extended up to $t_{\rm sweep}/N^2=200$, showed very slow relaxation: the block-resolved MSDs remained well below the Gaussian Rouse-like trends, while the junction monomer displayed only a small initial drift before reaching an apparent plateau. This near-arrest of the junction monomer is likely a consequence of the strong local constraints imposed by self-avoidance and its simultaneous coupling to both blocks. No common clean diffusive window could be identified from which a reliable $D_{\rm cm}(N)$ scaling could be extracted. 

The present model is best viewed as a minimal test case for rate-induced mobility heterogeneity in a Gaussian 2D lattice polymer. The local move geometry is deliberately kept fixed, and heterogeneity enters only through the relative frequency of attempted updates. This separates rate heterogeneity from other possible modifications, such as changing the step length, allowing enlarged local moves, or assigning different move dictionaries to different parts of the chain. Those variants may also be interesting, but they would mix mobility heterogeneity with changes in the local geometry of the dynamics. Related questions arise in heterogeneous Rouse models and partially active polymers~\cite{hung2018,osmanovic2017,osmanovic2018,vatin2024}.

A natural next step would be to connect this rate-based construction to more physical forms of heterogeneity. One example is a polymer segment moving in a random or dynamically heterogeneous environment, where local mobility varies from region to region; neutron-scattering studies of polymer blends have shown that chain dynamics can require a distribution of heterogeneous mobilities rather than a single average environment~\cite{niedzwiedz2007}. Another example is a crowded medium with mobile and immobile obstacles, where dynamic heterogeneity can strongly reduce diffusivity and modify polymer loop-formation kinetics~\cite{kwon2019}. In future extensions of the present model, one block could represent a region coupled to a more active or higher-temperature environment, while another could represent a segment with stronger steric constraints or reduced mobility in a crowded medium.

\section*{Acknowledgements}

I thank Jean-Charles Walter, Andrea Parmeggiani, and Linda Delimi of the Laboratoire Charles Coulomb, UMR 5221, CNRS--Universit\'{e} de Montpellier, for their guidance, encouragement, and insightful discussions on polymer physics, which helped shape the ideas leading to this work.

\section*{Code Availability}

The simulation code, processed data, and generated figures associated with this work are publicly available on GitHub at
\href{https://github.com/arpand2004/mobility-heterogeneity-2d-gaussian-polymer}{\texttt{https://github.com/arpand2004/mobility-heterogeneity-2d-gaussian-polymer}}
and archived on Zenodo at
\href{https://doi.org/10.5281/zenodo.20385055}{\texttt{https://doi.org/10.5281/zenodo.20385055}}.

\section*{Funding and Conflict of Interest Statement}

This work was carried out independently and did not receive any external funding. The author declares no competing financial or non-financial interests.

\end{document}